# Stability conditions for one-dimensional optical solitons in cubic-quintic-septimal media

Albert S. Reyna[1]*, Boris A. Malomed[2], and Cid B. de Araújo[1]

[1]*Departamento de Física, Universidade Federal de Pernambuco, 50670-901, Recife, PE, Brazil*

[2]*Department of Physical Electronics, School of Electrical Engineering, Faculty of Engineering, Tel Aviv University, Tel Aviv 69978, Israel*

- Corresponding author. E-mail: areynao@yahoo.com.br

## Abstract

Conditions for stable propagation of one-dimensional bright spatial solitons in media exhibiting optical nonlinearities up to the seventh-order are investigated. The results show well-defined stability regions even when all the nonlinear terms are focusing. Conditions for onset of the supercritical collapse of the optical beam are identified too. A variational approximation (VA) is used to predict dependence of the soliton's propagation constant on the norm, and respective stability regions are identified using the Vakhitov-Kolokolov (VK) criterion. Analytical results obtained by means of the VA are corroborated by numerical simulations of the cubic-quintic-septimal nonlinear Schrödinger equation.

## I.   Introduction

Optical spatial solitons are self-trapped light beams whose shape and transverse dimension remain invariant in the course of the propagation, due to the balance between diffraction and nonlinearity [1]. Different physical mechanisms may contribute to the generation of spatial solitons in nonlinear (NL) media [2]. Currently, the studies of spatial solitons constitute a very active field, with a potential for applications to photonics and optical communications [2-5]. In particular, it is well established that focusing Kerr-type media support the stable propagation of bright solitons in one transverse dimension [(1+1)D] [6]. Usually, the soliton's dynamics is described by the cubic nonlinear Schrödinger equation (C-NLSE), which gives rise to the currently known stable solution with the hyperbolic-secant envelope shape. Unstable soliton propagation is observed in (1+1)D when the system exhibits higher-order nonlinearities (HON) [7]. For example, in focusing quintic NL media the diffraction effect is not sufficient to balance the self-focusing, consequently the beam is subject to the critical collapse [8]. However, the inclusion of higher-order dissipative terms can suppress the collapse. In two transverse dimensions, [(2+1)D], the stationary soliton solutions of the C-NLSE (*Townes' solitons*) are extremely unstable against the propagation [9]. Therefore they are not observed in the usual Kerr media [10], although stable propagation can be observed when refractive and/or dissipative HON are considered.

Large HON were reported in various physical settings [11], playing an important role for the understanding of filamentation [12], harmonic conical diffraction [13], and other transverse NL phenomena [14-16]. In particular, as mentioned above, HON may help to stabilize the propagation of spatial solitons. The theoretical analysis shows that the formation of (1+1)D spatial solitons depends on the sign and magnitude of the third- and fifth-order NL terms [17]. In (2+1)D, stable soliton solution of the cubic-quintic NL Schrödinger equation (CQ-NLSE) was predicted by considering the competition between focusing third-order and defocusing fifth-order nonlinearities [18]. The experimental observation of stable (2+1)D fundamental solitons in cubic-quintic media, including a dissipative effect due to the three-photon absorption, was recently reported [19].

Several works were dealing with the stable propagation of solitons in media with cubic-quintic nonlinearities [20]. In addition to the cubic and quintic terms, septimal nonlinearity may also play an important role for the propagation of spatial solitons. Recently, (2+1)D bright spatial solitons were reported in metal colloids exhibiting quintic-septimal (focusing-defocusing) nonlinearity. Their behavior was modeled by the quintic-septimal NL Schrödinger equation (QS-NLSE) with dissipative terms [21].The nonlinearity management of NL systems, i.e., adjustment of the strength of the nonlinear terms, may allow controlled interplay between different NL terms, which leads to enhancement or suppression of specific HONs [14, 15, 21, 22]. In particular, a quintic medium with suppressed cubic nonlinearity was created in metal colloids by varying the volume fraction of silver nanoparticles (NPs) in acetone [14]. Based on a similar procedure of the nonlinearity management, a septimal medium was obtained by inducing a destructive interplay between cubic and quintic nonlinearities [15]. The availability of the management technique, which was experimentally demonstrated in Refs. [14, 15], justifies the theoretical effort to study the respective mathematical models.

A special case, which was studied, thus far, only theoretically is the critical collapse in focusing quintic media in the absence of the third-order nonlinearity. The corresponding model is provided by the quintic NL Schrödinger equation (Q-NLSE), which displays a degenerate family of 1D *Townes′ solitons* [23]. The addition of an external potential allows one to arrest the beam collapse in one [23] and two [24] transverse dimensions. In focusing cubic-quintic media, the cubic nonlinearity lifts the degeneracy, which is characteristic to the *Townes' solitons*, making the solitons' propagation stable against small perturbations [25]. Another case of interest is the study of the spatial-soliton propagation in septimal media. A highly unstable behavior, due to the high degree of the seventh-order nonlinearity, is expected, hence HON can give rise to a supercritical collapse of the beam. Thus, the study of cubic-quintic-septimal NL model is relevant to complement the previous studies, and it can suggest new possibilities for experiments.

In this work, conditions for the stable propagation of (1+1)D spatial solitons in media exhibiting nonlinearities up to the seventh order are analyzed. Section II presents the analytical model, based on the cubic-quintic-septimal NL Schrödinger equation (CQS-NLSE), for the description of the (1+1)D spatial-soliton propagation. A variational approximation (VA), based on the *raised-sech ansatz* [26], and the Vakhitov-Kolokolov (VK) stability criterion [27] are used to identify stability and instability regions of the soliton propagation. Section III demonstrates the evolution of stable and unstable fundamental solitons produced by numerical simulations of (1+1)D CQS-NLSE, which corroborates the analytical predictions reported in Section II. A summary of the results is presented in Section IV.

## II.   The analytical approximation for cubic-quintic-septimal model

The normalized (1+1)D CQS-NLSE, which describes the light propagation with one transverse dimension, in a medium exhibiting nonlinearities up to the seventh order, is

$$i\frac{\partial \psi}{\partial z} + \frac{1}{2}\frac{\partial^2 \psi}{\partial x^2} + g_3 \psi |\psi|^2 + g_5 \psi |\psi|^4 + \psi |\psi|^6 = 0, \tag{1}$$

where $\psi = \psi(x,z)$ is the normalized field amplitude, while $g_3$, $g_5$ and $g_7 \equiv +1$ represent the strength of the third, quintic and septimal NL terms, respectively. Dimensionless variables $x$ and $z$ are the transverse and propagation coordinates, respectively. The stability of the spatial solitons was studied by varying $g_3$ and $g_5$. It is important to clarify that the models in which the magnitudes of $g_3$, $g_5$ and $g_7$ are comparable, or, in some cases $g_7$ is dominant, do not violate the convergence principle of the power-series expansion. A clear example of that was observed experimentally in the NL behavior of metal-dielectric nanocomposites, and theoretically analyzed using the generalized Maxwell-Garnett model [14, 15].

Stationary solutions of Eq. (1) with a real propagation constant $k$ have the form of $\psi(x, z) = e^{ikz} \phi(x)$, with the real function $\phi = \phi(x)$ obeying the stationary equation:

$$0 = -k\phi + \frac{1}{2}\frac{d^2\phi}{dx^2} + g_3\phi^3 + g_5\phi^5 + \phi^7. \qquad (2)$$

An effective potential energy, $U$, can be defined by casting Eq. (2) into the form of $\partial^2\phi/\partial x^2 = -\partial U/\partial\phi$. Therefore, the corresponding Lagrangian density, $\mathcal{L} = (\phi')^2/2 - U$, is given by

$$\mathcal{L} = \frac{1}{2}(\phi')^2 + k\phi^2 - \frac{1}{2}g_3\phi^4 - \frac{1}{3}g_5\phi^6 - \frac{1}{4}\phi^8. \qquad (3)$$

As in the case of the Q-NLSE, Eq. (1) is nonintegrable, hence the VA is necessary to predict conditions for the stable soliton propagation. Taking into regard the commonly known fact that, for the ordinary cubic nonlinearity, the exact solution for the (1+1)D C-NLSE is $\phi(x) \propto \mathrm{sech}\left(\sqrt{2k}\,x\right)$, we here adopt the *raised-sech ansatz* [26],

$$\phi(x) = \Lambda\left[\mathrm{sech}\left(\sqrt{2k}\,x\right)\right]^\alpha, \qquad (4)$$

where $\Lambda$ and $\alpha$ are variational parameters. This *ansatz* allows one to control the beam's radius by changing the parameter $\alpha$.

The total power, $P = \int_{-\infty}^{\infty}\left[\phi(x)\right]^2 dx$, of *ansatz* (4) is

$$P = \frac{\Lambda^2}{\sqrt{2k}}\left[\frac{\sqrt{\pi}\,\Gamma(\alpha)}{\Gamma\left(\frac{1}{2}+\alpha\right)}\right], \qquad (5)$$

where $\Gamma$ is the Gamma-function. Then, by substituting the *raised-sech ansatz* into Eq. (3) and integrating over the 1D space, the following expression for the effective Lagrangian is obtained:

$$L = kP\left[1+\frac{\alpha^2}{2\alpha+1}\right] - g_3 P^2\sqrt{\frac{k}{2\pi}}\left[\frac{\Gamma(2\alpha)}{\Gamma\left(\frac{1}{2}+2\alpha\right)}\right]\left[\frac{\Gamma\left(\frac{1}{2}+\alpha\right)}{\Gamma(\alpha)}\right]^2$$

$$-\frac{2}{3\pi}g_5 kP^3\left[\frac{\Gamma(3\alpha)}{\Gamma\left(\frac{1}{2}+3\alpha\right)}\right]\left[\frac{\Gamma\left(\frac{1}{2}+\alpha\right)}{\Gamma(\alpha)}\right]^3 - \frac{1}{\sqrt{2}\pi^{\frac{3}{2}}}k^{\frac{3}{2}}P^4\left[\frac{\Gamma(4\alpha)}{\Gamma\left(\frac{1}{2}+4\alpha\right)}\right]\left[\frac{\Gamma\left(\frac{1}{2}+\alpha\right)}{\Gamma(\alpha)}\right]^4,$$

$$(6)$$

The respective Euler-Lagrange equations for the variational parameters, $\partial L/\partial P = 0$ and $\partial L/\partial \alpha = 0$, lead to the following system of equations:

$$0 = k\left[1+\frac{\alpha^2}{2\alpha+1}\right] - g_3 P\sqrt{\frac{2k}{\pi}}A_2 - \frac{2}{\pi}g_5 kP^2 A_3 - \frac{4}{\sqrt{2}\pi^{\frac{3}{2}}}k^{\frac{3}{2}}P^3 A_4, \qquad (7)$$

$$0 = 2\alpha kP\left[\frac{\alpha+1}{(2\alpha+1)^2}\right] - g_3 P^2\sqrt{\frac{k}{2\pi}}B_2 - \frac{2}{3\pi}g_5 kP^3 B_3 - \frac{1}{\sqrt{2}\pi^{\frac{3}{2}}}k^{\frac{3}{2}}P^4 B_4, \qquad (8)$$

with

$$A_m = \left[ \frac{\Gamma\left(\frac{1}{2}+\alpha\right)}{\Gamma(\alpha)} \right]^m \left[ \frac{\Gamma(m\alpha)}{\Gamma\left(\frac{1}{2}+m\alpha\right)} \right], \tag{9}$$

$$B_m = \frac{\left[ \Gamma\left(\frac{1}{2}+\alpha\right) \right]^{m-1}}{\left[ \Gamma(\alpha) \right]^m \Gamma\left(\frac{1}{2}+m\alpha\right)} \left\{ m\Gamma(m\alpha) \left[ \Gamma'\left(\frac{1}{2}+\alpha\right) - \frac{\Gamma\left(\frac{1}{2}+\alpha\right)\Gamma'(\alpha)}{\Gamma(\alpha)} \right] \right.$$
$$\left. + \Gamma\left(\frac{1}{2}+\alpha\right) \left[ \Gamma'(m\alpha) - \frac{\Gamma(m\alpha)\Gamma'\left(\frac{1}{2}+m\alpha\right)}{\Gamma\left(\frac{1}{2}+m\alpha\right)} \right] \right\}, \tag{10}$$

where $\Gamma'$ is the derivative of the Gamma-function. Due to their complexity, Eqs. (7) and (8) had to be solved numerically.

Figure 1 shows results obtained by numerical solution of the full stationary version of the CQS-NLSE, Eq. (2) (circles and triangles), and produced by the VA (dashed lines), i.e., by the numerical solution of Eqs.(7) and (8). The red line and circles illustrate the dependence of the propagation constant, $k$, on the soliton's power, $P$, for the septimal-only medium $\left( g_3 = 0, g_5 = 0 \right)$, while the blue line and the triangles display the same for a quintic-septimal medium $\left( g_3 = 0, g_5 \neq 0 \right)$. The soliton's stability regions were identified on the basis of the VK stability criterion [27], $\partial P/\partial k > 0$. For the two cases shown in Fig. 1 we observe $\partial P/\partial k < 0$, indicating that the spatial soliton is unstable in the septimal-only and quintic-septimal media. We recall that the quintic-only medium also gives rise to the instability [9]. In the present case, the addition of the positive seventh-order term leads to additional strong self-focusing, which may result in a supercritical collapse, as shown below.

By contrast, when the focusing cubic nonlinearity is also present $\left(g_3 > 0\right)$, regions of stability are observed for different values of $g_5$, as shown in Fig. 2, where the third- and seventh-order nonlinearities are fixed to be $g_3 = +1$ and $g_7 = +1$, respectively. In that figure, solid (dashed) lines correspond to regions of stable (unstable) soliton propagation, identified by $\partial P / \partial k > 0$ $\left(\partial P / \partial k < 0\right)$. Also, it is possible to observe that the maximum power, $P_{\max}$, for the stable soliton propagation in media with $g_5 = -1$ (red line and circles) is larger than in the medium with $g_5 = 0$ (blue line and triangles), and larger too than in the medium with $g_5 = +1$ (black line and squares). The negative (defocusing) fifth-order nonlinearity balances the self-focusing effect, enlarging the stability region, while the positive (focusing) fifth-order term accelerates the onset of the critical self-focusing.

To extend these conclusions, numerical solutions of Eq. (2), (7) and (8) were obtained for various values of $g_5$ between $-1.5$ and $+1.5$, with fixed $g_3 = +1$. For each value of $g_5$, the maximum power, $P_{\max}$, which allows the stable soliton propagation, was found. Figure 3 shows that, with the growth of the quintic nonlinearity (going from negative to positive, i.e., from defocusing to focusing), the stability region for the soliton propagation is reduced. Therefore, large negative values of $g_5$ help to stabilize the soliton, while large positive values of $g_5$ promote the onset of the collapse, even at low powers.

From the analysis of Figs. 1-3, we conclude that the VA using the *raised-sech ansatz* adequately describes the spatial soliton propagation in media exhibiting nonlinearities up to the seventh order.

### III.   Numerical simulations of (1+1)D cubic-quintic-septimal nonlinear Schrödinger equation

The beam propagation in the present model was simulated by solving numerically the full CQS-NLSE, Eq. (1), using the split-step compact finite-difference method [28]. Stability and instability regions, predicted by the VA with the help of the VK criterion, were verified by using the *raised-sech ansatz*, given by Eq. (4), as the input for the direct simulations.

Figure 4 exhibits the beam's collapse in the medium without third-order nonlinearity, after a very small propagation distance. Figure 4(a) corresponds to the beam propagation in the septimal-only medium $\left( g_3 = g_5 = 0 \right)$, for the soliton's propagation constant $k = 1$, with the respective values, $P = 1.51$ and $\alpha = 1.97$ obtained from Eqs. (7) and (8). Strong self-focusing is observed at $z \approx 1.3$, resulting in the formation of jets induced by the seventh-order nonlinearity (the *supercritical collapse*). A still faster collapse is observed by adding the focusing quintic term $\left( g_3 = 0, g_5 = +1 \right)$, which additionally contributes to the development of the collapse, as shown in Fig. 4(b). The values of $k = 1$, $P = 1.23$ and $\alpha = 1.90$, predicted by the VA, were used to construct the input beam for the simulations of the quintic-septimal medium. In both cases, unstable propagation of spatial solitons is observed due to the dominant role of the septimal nonlinearity, in agreement with Fig. 1.

Figure 5 shows the predicted beam propagation in the cubic-quintic-septimal medium, for different values of $k$. Figures 5(a)-5(d) exhibit a stable family of solitons obtained by solving the CQS-NLSE with $g_3 = +1$ and $g_5 = +1$. For values of $k$ and $P$ below the limit values for the stable propagation $\left( k_{\max}, P_{\max} \right) = \left( 1.19, 1.11 \right)$, it is possible to observe the formation of periodically oscillating breathers, probably due to a small inaccuracy of the input with respect to the exact waveform. On the other hand, Figs. 5(e) and 5(f) show the collapse for ($k=1.5$, $P=1.11$) and ($k=2.0$, $P=1.10$), respectively. Thus, the stability boundaries, predicted by the VA in the combination with the VK criterion, enable the identification of the stability boundaries, which separate the formation of the fundamental soliton and the collapse, with good accuracy.

Additional simulations of Eq. (1) were performed to confirm the predictions of the VA. In particular, similar results were obtained for the media with $\left(g_3=+1, g_5=-1\right)$ and $\left(g_3=+1, g_5=0\right)$, in agreement with Fig. 2.

From the experimental point of view, suitable conditions for observing stability regions, as well as the critical and supercritical collapse, may be provided by the nonlinearity management procedure reported in Refs. [14, 15, 21]. In particular, HONs were observed with peak powers of few kW, using picoseconds pulses at 532 nm in colloids with volume fractions of silver nanoparticles in the range of $10^{-5}$ to $10^{-4}$. Experiments in the infrared will help to expand the relevant parameter space, as one may flip the signs of the nonlinear refractive indices of different orders by varying the detuning with respect to the surface-plasmon resonance in the nanoparticles. Thus, taking into regard nonlinearity parameters reported in Refs. [14, 15, 21] and perspective of further experiments with different laser wavelengths and pulse durations, it should be quite realistic to reach conditions for the observation of the effects predicted in the present work, using planar waveguides filled with silver colloids.

**IV.   Summary**

We have reported the detailed analysis of conditions for the stable propagation of (1+1)D spatial solitons in media exhibiting nonlinearities up to the seventh order. Stability and instability regions were identified using the combination of the VA (variational approximation) and VK (Vakhitov-Kolokolov) criterion. Regions of stable soliton propagation were identified for the media exhibiting the focusing cubic and septimal terms, with either sign of the quintic term. The analytical results were verified by numerical simulations of the full underlying NLSE with the CQS (cubic-quintic-septimal) nonlinearity, which show close agreement with the predictions of the VA. The direct simulations corroborate that the unstable solitons suffer the catastrophic self-focusing, as it might be naturally expected.

**Acknowledgments**

We acknowledge financial support from the Brazilian agencies Conselho Nacional de Desenvolvimento Cientifico e Tecnológico (CNPq) and Fundação de Amparo à Ciência e Tecnologia do Estado de Pernambuco (FACEPE). The work was performed in the framework of the National Institute of Photonics (INCT de Fotônica) project and PRONEX/CNPq/FACEPE. B. A. M. appreciates hospitality of the Department of Physics at the Universidade Federal de Pernambuco (Recife, Brazil).

**REFERENCES**

[1] R. Y. Chiao, E. Garmire, and C. H. Townes, Phys. Rev. Lett. **13**, 479 (1964); G. I. Stegeman and M. Segev, Science **286**, 1518 (1999); Y. Kivshar, Nature Phys. **2**, 729 (2006).

[2] Z. Chen, M. Segev, and D. N. Christodoulides, Rep. Prog. Phys. **75**, 086401 (2012). Y. V. Kartashov, B. A. Malomed, and L. Torner, Rev. Mod. Phys. **83**, 247 (2011).

[3] Y. S. Kivshar and G. P. Agrawal, Optical Solitons: From Fibers to Photonic Crystals (Academic, 2003).

[4] M. Tiemann, T. Halfmann, and T. Tschudi, Opt. Commun. **282**, 3612 (2009).

[5] A. V. Buryak, P. Di Trapani, D. V. Skryabin, and S. Trillo, Phys. Rep. **370**, 63 (2002).

[6] A. Barthelemy, S. Maneuf, and C. Froehly, Opt. Commun. **55**, 201 (1985); J. S. Aitchison, A. M. Weiner, Y. Silberberg, M. K. Oliver, J. L. Jackel, D. E. Leaird, E. M. Vogel, and P. W. E. Smith, Opt. Lett. **15**, 471 (1990).

[7] O. Bang, J. J. Rasmussen, and P. L. Christiansen, Nonlinearity **7**, 205 (1994).

[8] Y. Chung and P. M. Lushnikov, Phys. Rev. E **84**, 036602 (2011).

[9] J. J. Rasmussen and K. Rypdal, Phys. Scr. **33**, 481 (1986).

[10] P. L. Kelley, Phys. Rev. Lett. **15**, 1005 (1965); E. L. Dawes and J. H. Marburger, Phys. Rev. **179**, 862 (1969); J. H. Marburger, Prog. Quantum Electron. **4**, 35 (1975).

[11] V. Loriot, E. Hertz, O. Faucher, and B. Lavorel, Opt. Express **17**, 13429 (2009); C. Schnebelin, C. Cassagne, C. B. de Araújo, and G. Boudebs, Opt. Lett. **39**, 5046 (2014); E. L. Falcão-Filho, C. B. de Araújo, and J. J. Rodrigues Jr., J. Opt. Soc. Am. B **24**, 2948 (2007); K. Dolgaleva, H. Shin, and R. W. Boyd, Phys. Rev. Lett. **103**, 113902 (2009); M. S. Zubairy, A. B. Matsko, and M. O. Scully, Phys. Rev. A **65**, 043804 (2002); F.

Smektala, C. Quémard, V. Couderc and A. Barthélémy, J. Non-Cryst. Solids **274**, 232 (2000).

[12] C. Brée, A. Demircan, and G. Steinmeyer, Phys. Rev. Lett. **106**, 183902 (2011); P. Béjot, J. Kasparian, S. Henin, V. Loriot, T. Vieillard, E. Hertz, O. Faucher, B. Lavorel, and J.-P. Wolf, Phys. Rev. Lett. **104**, 103903 (2010); G. Point, Y. Brelet, A. Houard, V. Jukna, C. Milián, J. Carbonnel, Y. Liu, A. Couairon, and A. Mysyrowicz, Phys. Rev. Lett. **112**, 223902 (2014).

[13] K. D. Moll, D. Homoelle, A. L. Gaeta, and R. W. Boyd, Phys. Rev. Lett. **88**, 153901 (2002); D. L. Weerawarne, X. Gao, A. L. Gaeta, and B. Shim, Phys. Rev. Lett. **114**, 093901 (2015).

[14] A. S. Reyna and C. B. de Araújo, Phys. Rev. A **89**, 063803 (2014).

[15] A. S. Reyna and C. B. de Araújo, Opt. Express **22**, 22456 (2014).

[16] M. Saha and A. K. Sarma, Opt. Commun. **291**, 321 (2013).

[17] Kh. I. Pushkarov, D. I. Pushkarov, and I. V. Tomov, Opt. Quantum Electron. **11**, 471 (1979); F. Kh. Abdullaev and J. Garnier, Phys. Rev. E **72**, 035603 (2005).

[18] V. Skarka, V. I. Berezhiani, and R. Miklaszewski, Phys. Rev. E **56**, 1080 (1997).

[19] E. L. Falcão-Filho, C. B. de Araújo, G. Boudebs, H. Leblond, and V. Skarka, Phys. Rev. Lett. **110**, 013901 (2013).

[20] N. VietHung, M. Trippenbach, E. Infeld, and B. A. Malomed, Phys. Rev. A **90**, 023841 (2014); R. M. Caplan, R. Carretero-González, P. G. Kevrekidis, and B. A. Malomed, Math. Comput. Simul. **82**, 1150 (2012); C. Rogers, B. Malomed, J. H. Li, and K. W. Chow, J. Phys. Soc. Jpn. **81**, 094005 (2012); J. Belmonte-Beitia and J. Cuevas, J. Phys. A: Math. Theor. **42**, 165201 (2009); C. Chong, R. Carretero-González, B. A. Malomed, and P. G. Kevrekidis, Physica D **238**, 126 (2009).

[21]  A. S. Reyna, K. C. Jorge, and C. B. de Araújo, Phys. Rev. A **90**, 063835 (2014).

[22]  M. Centurion, M. A. Porter, P. G. Kevrekidis, and D. Psaltis, Phys. Rev. Lett. **97**, 033903 (2006); F. Ö. Ilday and F. W. Wise, J. Opt. Soc. Am. B **19**, 470 (2002); R. Driben, B. A. Malomed, and U. Mahlab, Opt. Commun. **232**, 129 (2004); A. Radosavljević, G. Gligorić, A. Maluckov, M. Stepić, and D. Milović, J. Opt. Soc. Am. B **30**, 2340 (2013); P. A. Subha, C. P. Jisha, and V. C. Kuriakose, Pramana **69**, 229 (2007); J. Fujioka, E. Cortés, R. Pérez-Pascual, R. F. Rodríguez, A. Espinosa, and B. A. Malomed, Chaos **21**, 033120 (2011).

[23]  Y. B. Gaididei, J. Schjodt-Eriksen, and P. L. Christiansen, Phys. Rev. E **60**, 4877 (1999); J. Zeng and B. A. Malomed, Phys. Rev. A **85**, 023824 (2012); G. L. Alfimov, V. V. Konotop, and P. Pacciani, Phys. Rev. A **75**, 023624 (2007).

[24]  B. B. Baizakov, B. A. Malomed, and M. Salerno, Europhys. Lett. **63**, 642 (2003).

[25]  D. E. Pelinovsky, Y. S. Kivshar, and V. V. Afanasjev, Physica D **116**, 121 (1998).

[26]  J. Zeng and B. A. Malomed, Phys. Rev. E **86**, 036607 (2012).

[27]  N. G. Vakhitov and A. A. Kolokolov, Radiophys. Quantum Electron. **16**, 783 (1973).

[28]  S. Wang and L. Zhang, Comput. Phys. Comm. **184**, 1511 (2013).

**Figure captions**

1. (Color online) The soliton's propagation constant, $k$, versus the total power , $P$, for media with suppressed third-order nonlinearity $(g_3 = 0)$. Discrete points correspond to the solution of Eq. (2) with $g_5 = 0$ (circles) and $g_5 = 1$ (triangles). Dashed lines were obtained using the variational approximation.

2. (Color online) Dependence of the soliton's propagation constant, $k$, on the total power, $P$, obtained from solution of Eq. (2) and from the variational approximation for media with $g_3 = 1$ and $g_5 = -1$ (red line and circles), $g_5 = 0$ (blue line and triangles) and $g_5 = 1$ (black line and squares). Solid (dashed) lines represent stable (unstable) solitons, as per the VK criterion.

3. (Color online) The maximum power, $P_{\max}$, admitting the stable soliton propagation in media, as per the VK criterion, with the focusing cubic nonlinearity $(g_3 = 1)$ and different values of the quintic coefficient $g_5$.

4. (Color online) The supercritical collapse of the beam produced by numerical solutions of the CQS-NLSE in media with the third-order nonlinearity suppressed $(g_3 = 0)$ and (a) $g_5 = 0$; (b) $g_5 = 1$. Here and in Fig. 5, the region after the onset of the collapse is not displayed, as simulations of the present model are not sufficient for tracing the post-collapse evolution.

5. (Color online) The evolution of stable and unstable fundamental solitons in the focusing cubic-quintic-septimal media $(g_3 = 1, g_5 = 1)$ with values of $(k, P, \alpha)$ taken as (a)

$\left(10^{-4}, 0.05, 1.47\right)$, (b) $\left(0.05, 0.63, 1.48\right)$, (c) $\left(0.3, 1.0, 1.60\right)$, (d) $\left(1, 1.11, 1.69\right)$, (e)

$\left(1.5, 1.11, 1.70\right)$ and (f) $\left(2, 1.10, 1.77\right)$.

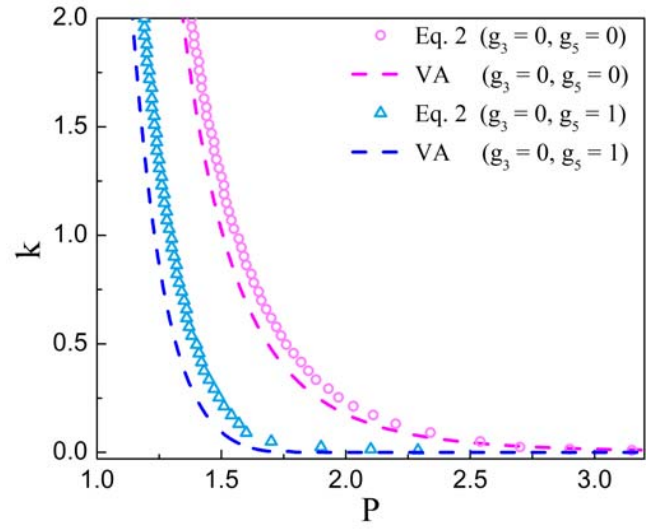

Fig. 1 Reyna et al.

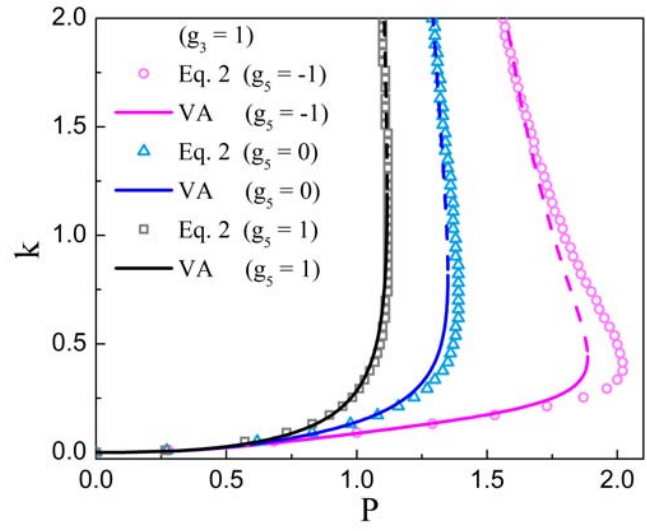

Fig. 2 Reyna et al.

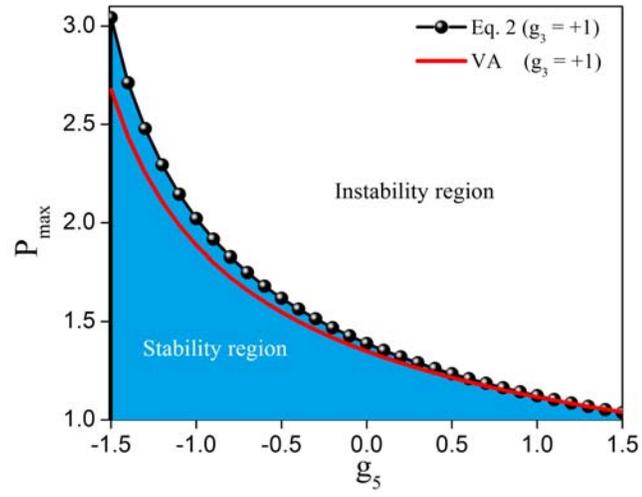

Fig. 3 Reyna et al.

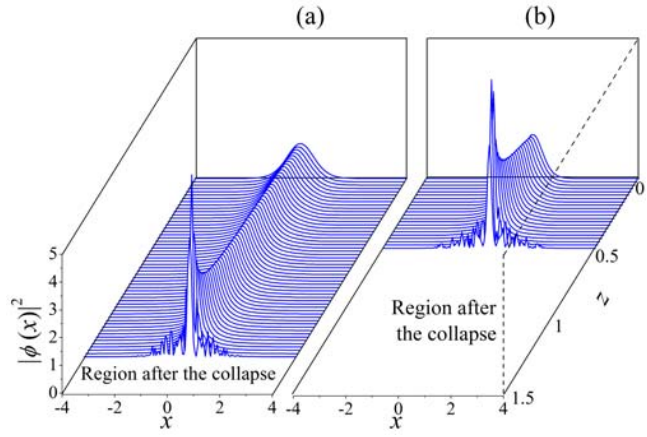

Fig. 4 Reyna et al.

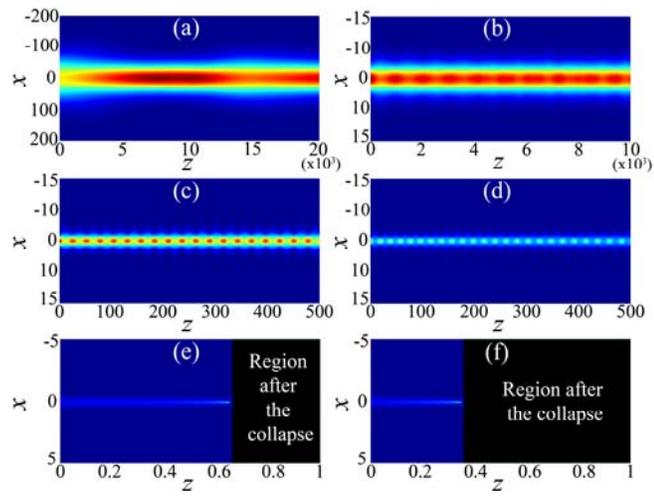

Fig. 5 Reyna et al.